# An accelerator scenario for hard X-ray free electron laser joint with high energy electron radiography

Tao Wei(魏涛) [1,2;1], Yiding Li(李一丁) [1], Guojun Yang(杨国君) [1], Jian Pang(庞健) [1], Yuhui Li(李煜辉) [2], Peng Li(李鹏) [3,2], Joachim Pflueger [2], Xiaozhong He(何小中) [1], Yaxing Lu(卢亚鑫) [1], Ke Wang(王科) [1], Jidong Long(龙继东) [1], Linwen Zhang(章林文) [1], Qiang Wu(吴强) [1]

[1] Institute of Fluid Physics, China Academy of Engineering Physics, P.O. Box 919-106, Mianyang 621900, China

[2] European XFEL Gmbh, Hamburg 22607, Germany

[3] Institute of Applied Electronics, China Academy of Engineering Physics, Mianyang 621900, China

Abstract: In order to study the dynamic response of the material and the physical mechanism of the fluid dynamics, an accelerator scenario which can be applied to hard X-ray free electron laser and high energy electron radiography was proposed. This accelerator is mainly composed of a 12GeV linac, an undulator branch and an eRad beamline. In order to characterize sample's dynamic behavior in situ and real-time with XFEL and eRad simultaneously, the linac should be capable of accelerating the two kinds of beam within the same operation mode. Combining with in-vacuum and tapering techniques, the undulator branch can produce more than $10^{11}$ photons per pulse in 0.1% bandwidth at 42keV. Finally, the eRad amplifying beamline with 1:10 ratio was proposed as an important complementary tool for the wider view field and density identification ability.

## 1. Introduction

In order to study the dynamic response of the material and the physical mechanism of the fluid dynamics, it is suggested that there should be simultaneously three probe scales [1], the micro level, meso level and macro level on which the detection of multi-spatial and multi-temporal phenomena can be made.

At macro scale, high energy proton radiography [2,3] (pRad) technique is a capable diagnostic tool which can detect the samples up to decimeter view field with the maximum resolution of tens microns. It is a pity that pRad is hard to achieve higher resolution even for the thinner samples because of the beam quality of proton beam.

In comparison with proton beam, the electron beam can easily obtain better quality because of less impact caused by space charge effect. Thus high energy electron radiography [4] (eRad) could have better performance at meso scale for the advantage of less emittance and energy spread. It is worth noting that eRad cannot replace pRad at macro scale because of the poor penetrating power.

At micro scale, X-ray free electron laser (XFEL) is a unique diagnostic tool which can achieve the resolution ability from micron to sub-nanometer [5].

## 2. Linac

In order to produce high brightness XFEL, an electron linear accelerator (linac) which acts as beam provider is indispensible. There is no doubt that such a linac can also provide beam for eRad. However the requirements are quite different between XFEL beam and eRad beam, the key point is to ensure the simultaneous acceleration of the two kinds of electron bunches within the same operation mode. As shown in Tab.1 is the beam parameters for XFEL and eRad individually.

* Supported by China Academy of Engineering Physics (2014A0402016) and Institute of Fluid Physics (SFZ20140201)
1) E-mail: tao.wei@desy.de

Tab.1. Parameters of XFEL beam and eRad beam at the end of linac

|  | XFEL electron bunch | eRad electron bunch |
|---|---|---|
| Beam energy (GeV) | 12 | 12 |
| Bunch charge (nC) | 0.2 | 1 |
| Rms bunch length (fs) | 22.5 | 230 |
| Peak current (kA) | 3 | 1.5 |
| Normalized slice emittance (μm) | 0.25 | 1.08 |
| Normalized emittance (μm) | 0.32 | 1.1 |
| Energy spread (%) | 0.015 | 0.2 |

As shown in Fig.1 is the sketch map of this accelerator, two kinds of electron bunches which used for XFEL and eRad individually are produced by the injector, all of them are accelerated to 12GeV energy at the end of linac, and then they are separated into different branches. The XFEL bunches move along the sinusoidal trajectory under the periodic magnetic field from undulator and emit coherent electromagnetic radiation which is called free electron laser. After shaping, focusing and monochromatization in the X-ray transport line, the XFEL arrive the sample eventually. The eRad bunches are kicked into another branch, after a longer trip they pass through the sample and come into being a reversed image at the end of the eRad beamline.

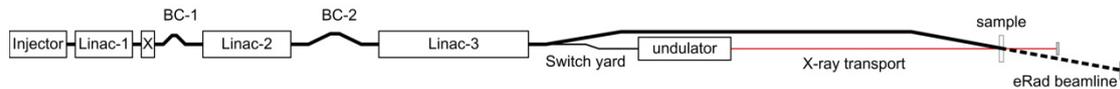

Fig.1. schematically diagram of the accelerator for XFEL and eRad simultaneously

The performance of XFEL and eRad are directly dependent on the beam quality of the electron bunches as produced by the injector and preserved by the main linac. In order to acquire a sufficiently small emittance and energy spread, the photocathode injector with two drive laser system was proposed. The injector can provide two kinds of electron bunches (see Tab.2).

Tab.2. Parameters of XFEL beam and eRad beam at the end of injector

|  | XFEL electron bunch | eRad electron bunch |
|---|---|---|
| Beam energy (MeV) | 150 | 150 |
| Bunch charge (nC) | 0.2 | 1 |
| Rms bunch length (ps) | 2 | 3 |
| Peak current (A) | 30 | 100 |
| Normalized slice emittance (μm) | 0.20 | 1.0 |
| Incoherent energy spread (%) | 0.008 | 0.01 |

In order to reach SASE saturation in a reasonable length undulator, the designed XFEL

electron beam should have a peak current of 3kA. Such a high current density cannot be achieved directly by injector under the premise of beam quality. Therefore the XFEL electron bunch is accelerated and compressed in a series of linacs (Linac-1, Linac-2 and Linac-3) and magnetic chicanes (BC-1 and BC-2) as shown in Fig.1. Extensive simulation studies from elegant [6] code indicate the beam quality is maintained very well.

The nominal linac design parameters for 0.2nC XFEL electron beam are summarized numerically in Tab.3 and graphically in Fig.2 below. Linac-1 is composed of three 3m long S-band accelerating cavities ($f_{rf}$=2.856GHz), it accelerates the XFEL electron bunch to 285MeV with off-crest phase and generates energy-time correlation along the bunch as shown in Fig.2(a). The energy-time correlation which generally called energy chirp is initial preparing for the bunch compression. A short high order harmonic cavity (0.6m long, $f_{rf}$=11.424GHz) is designed to remove the quadratic energy-time correlation. As shown in Fig.2(b), Linac-X can cancel rf curvature non-linearity generated in Linac-1, it operates at 180° and decelerates the bunch energy by 12MeV.

Tab.3. Nominal linac parameters for 0.2nC XFEL electron bunch

|  | $E_{in}$ (GeV) | $E_{out}$ (GeV) | $\sigma_{t\text{-}in}$ (ps) | $\sigma_{t\text{-}out}$ (ps) | $\sigma_{\delta\text{-}in}$ (%) | $\sigma_{\delta\text{-}out}$ (%) | $\varphi_{rf}$ (deg) | $R_{56}$ (mm) |
|---|---|---|---|---|---|---|---|---|
| Linac-1 | 0.15 | 0.28 | 2.0 | 2.0 | 0.01 | 1.1 | -33.5 | — |
| Linac-X | 0.28 | 0.27 | 2.0 | 2.0 | 1.1 | 1.15 | 180 | — |
| BC-1 | 0.27 | 0.27 | 2.0 | 0.26 | 1.15 | 1.15 | — | -44.9 |
| Linac-2 | 0.27 | 2.08 | 0.26 | 0.26 | 1.15 | 0.4 | -37.5 | — |
| BC-2 | 2.08 | 2.08 | 0.26 | 0.023 | 0.4 | 0.4 | — | -17.9 |
| Linac-3 | 2.08 | 12.18 | 0.023 | 0.023 | 0.4 | 0.015 | +12 | — |

The first bunch compressor [7] is a 4-dipole chicane (BC-1), which shortens the XFEL electron bunch to 260fs rms. Due to the influence of coherent synchrotron radiation [8] (CSR) and longitudinal space charge effect (LSC), the incoherent energy spread will increase as shown in Fig.2(c). In fact, the parameter selection of bunch compressor is always a compromise between beam quality and compression ratio. Linac-2 ($f_{rf}$=2.856GHz) accelerates the bunch to 2.08GeV, also at an off-crest phase, which maintains the desired linear energy chirp as shown in Fig.2(d).

The second bunch compressor is another 4-dipole chicane (BC-2) which compresses the XFEL electron bunch to its final value of 22.5fs as shown in Fig.2(e). Finally, Linac-3 accelerates the bunch to 12GeV, in which positive acceleration phase is chosen to cancel the time-correlated coherent energy spread.

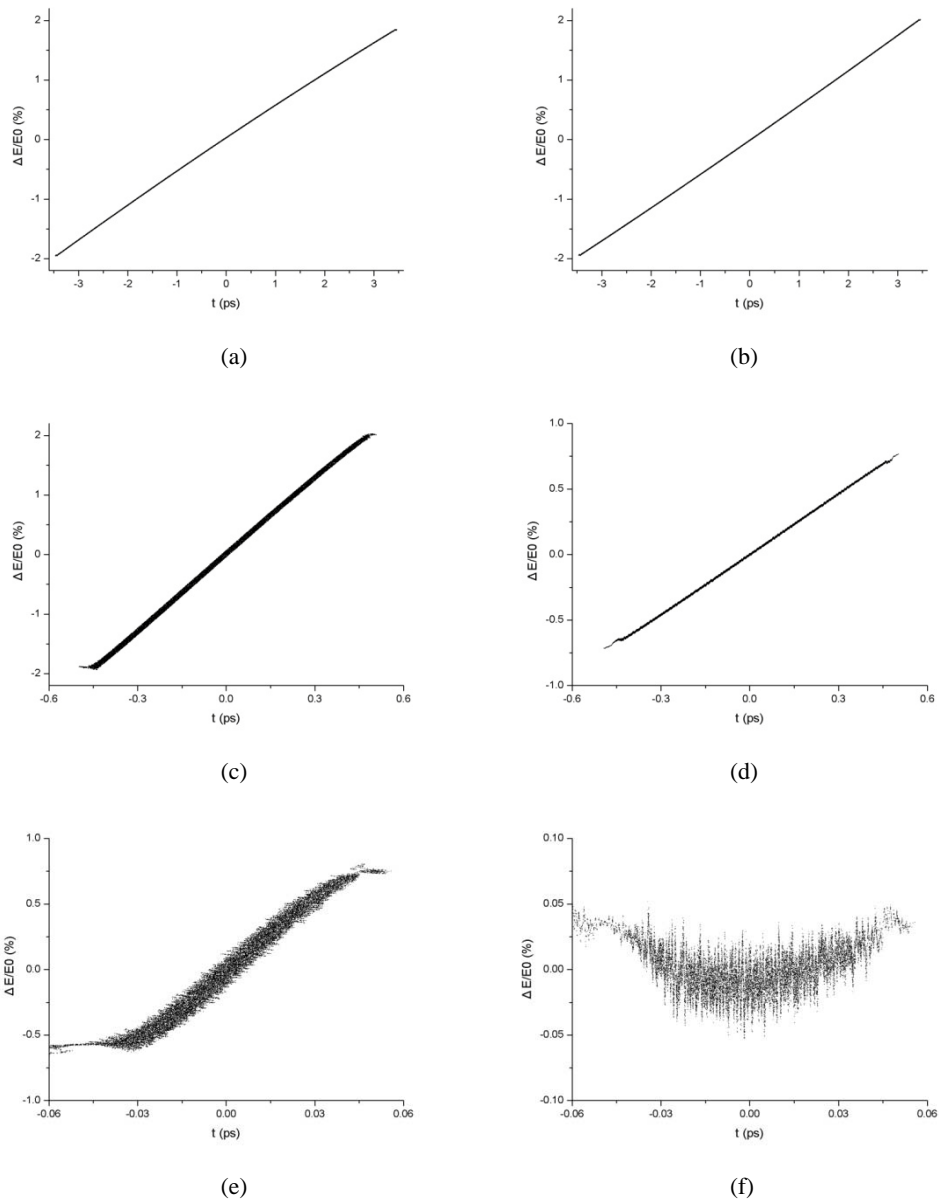

Fig.2. The Longitudinal phase space distribution of 0.2nC XFEL electron bunch after Linac-1 at 0.28GeV (figure a), after Linac-X at 0.27GeV (figure b), after BC-1 7.7 times compression section (figure c), after Linac-2 at 2.08GeV (figure d), after BC-2 11 times compression section (figure e) and after Linac-3 at 12.17GeV (figure f).

As shown in Fig.3 is the current distribution at the end of linac. The very high peak current generated by the compression process can drive micro-bunching instability, which can damage both the slice emittance and the slice energy spread. In order to damp this effect, a laser-heater is placed in the injector section to adjust the initial incoherent energy spread. The simulation result shows $8\times10^{-5}$ initial incoherent energy spread is sufficient to suppress the instability without adding too much energy spread at the entrance of undulator.

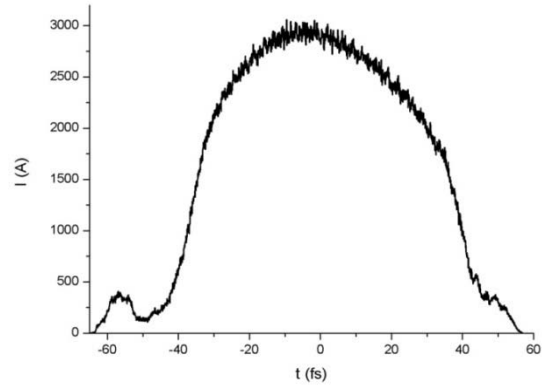

Fig.3. The current distribution of 0.2nC XFEL electron bunch along the beam length at the end of linac

All mentioned above are the longitudinal dynamics of 0.2nC XFEL electron bunch, in fact the transverse dynamics are also necessary, especially for the beam emittance.

A FODO structure with betatron phase advance 60° per cell is chosen as the focusing lattice for Linac-1. Three acceleration cavities are located in 1.5 cells with equally spaced 0.4m insert section where 0.1m long quadrupole magnet and BPM-steering pairs can be arranged. A short X-band cavity is located in the following drift. BC-1 is composed of four 0.2m long sector dipole magnets, it is an achromatic structure naturally and the dispersion function remains zero in the outer space. In order to weaken CSR effect in BC-1 section, horizontal beta function ($\beta_x$) should be as small as possible, especially in the latter two dipoles when the bunch compression reaches the maximum value. As shown in Fig.4, $\beta_x$ is small enough in BC-1 and the emittance dilution due to CSR effect is less than 10%.

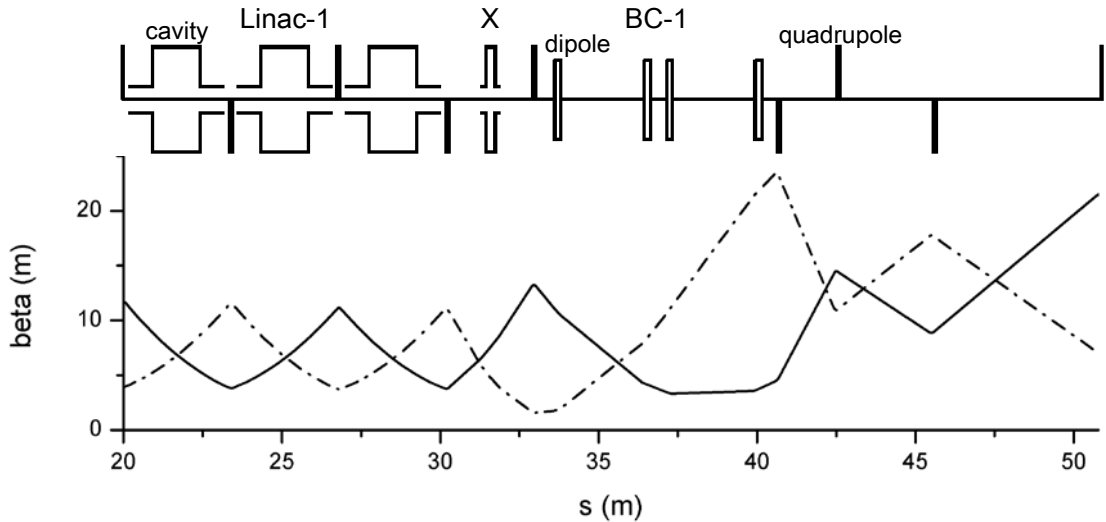

Fig.4. Twiss parameters along Linac-1 & Linac-X (S<33m) and through BC-1 chicane and emittance diagnostic section (S>41m), the solid line represents $\beta_x$ and the dash line represents $\beta_y$.

A FODO structure with 60° phase advance per cell and 13m period length is used in Linac-2. Every two cavities are connected together and placed between the quadrupole magnets as shown in Fig.5.

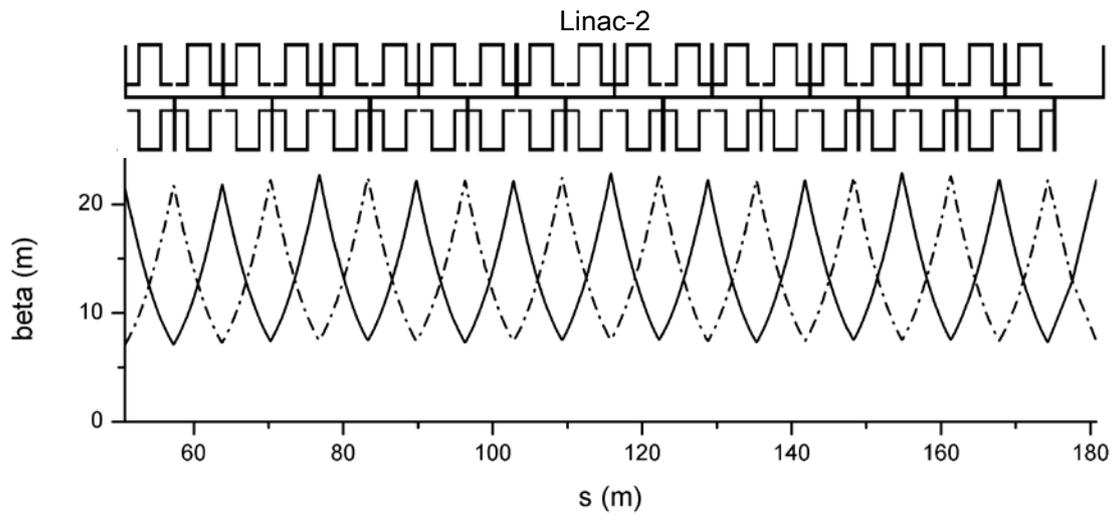

Fig.5. Twiss parameters along Linac-2, in which the solid line represents $\beta_x$ and the dash line represents $\beta_y$.

BC-2 compression section has nearly the same design discipline as BC-1 except the stronger CSR effect. Moreover, the beam diagnostic sections are necessary at both ends. The emittance dilution can be controlled within 20% by selecting appropriate twiss parameters as Fig.6.

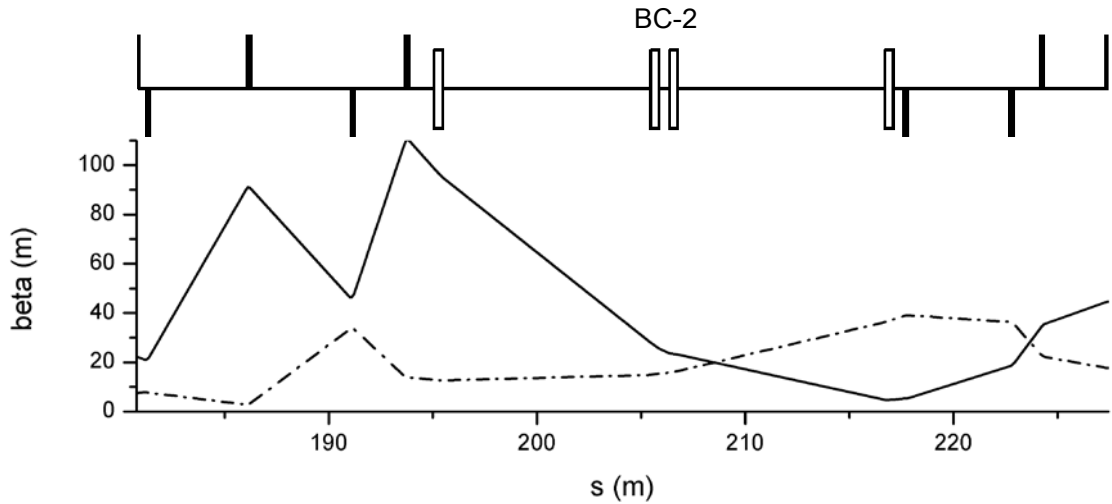

Fig.6. Twiss parameters along BC-2 chicane and beam diagnostic sections (S<194m and S>218m), the solid line represents $\beta_x$ and the dash line represents $\beta_y$.

Linac-3 adopts the FODO structure with 54° phase advance per cell and 25.2m period length, as shown in Fig.7, every four cavities connected together and 168 cavities located in 21 cells. In order to provide enough focusing strength, the quadrupole magnets are up to 0.2m long.

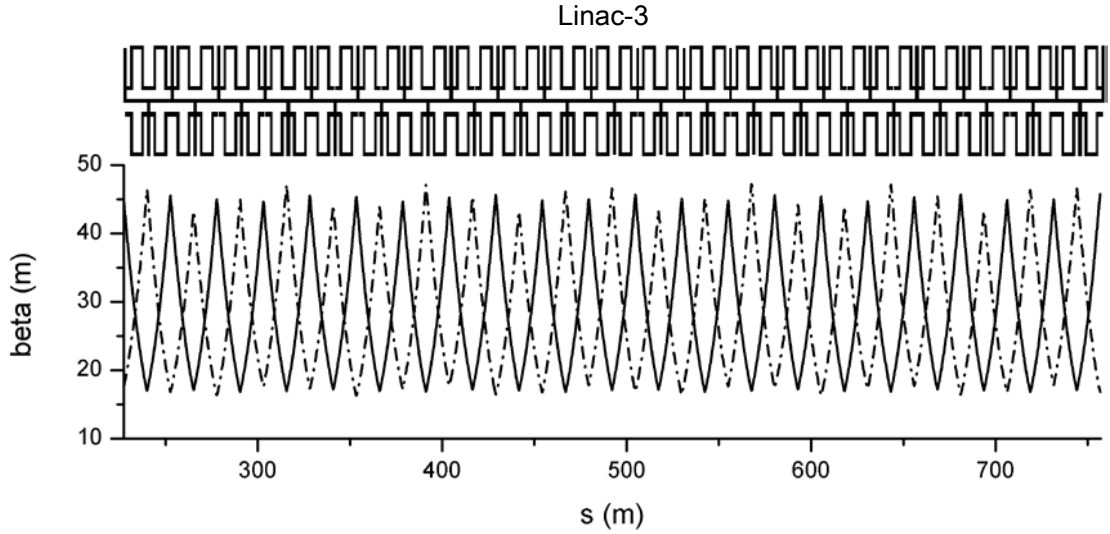

Fig.7. Twiss parameters along Linac-3, the solid line represents $\beta_x$ and the dash line represents $\beta_y$.

All the parameters mentioned above are suitable for 0.2nC XFEL electron bunch, and the beam quality can satisfy the design guideline by means of these parameters. But these parameters are not capable for 1nC eRad electron bunch because it suffers much more serious instability than the former.

In order to achieve the simultaneous acceleration of two kinds of beam, all machine parameters remain unchanged except the initial phase of eRad beam is delayed 10° which can be implemented by adjusting the timing trigger of injector laser system. As shown in Tab.4, it is possible to completely change the linac parameters only through the variation of initial acceleration phase. By comparison Tab.3 and Tab.4, eRad electron bunch has less compression ratio and peak current, the corresponding emittance dilution can be ignored. There is a fly in the ointment that the coherent energy spread is hard to be cancelled totally. Even so, the beam quality of eRad electron bunch is still within specification.

Tab.4. Nominal linac parameters for 1nC eRad electron bunch

|  | $E_{in}$ (GeV) | $E_{out}$ (GeV) | $\sigma_{t\text{-}in}$ (ps) | $\sigma_{t\text{-}out}$ (ps) | $\sigma_{\delta\text{-}in}$ (%) | $\sigma_{\delta\text{-}out}$ (%) | $\varphi_{rf}$ (deg) | $R_{56}$ (mm) |
|---|---|---|---|---|---|---|---|---|
| Linac-1 | 0.15 | 0.3 | 3.0 | 3.0 | 0.02 | 1.16 | -23.5 | — |
| Linac-X | 0.3 | 0.29 | 3.0 | 3.0 | 1.16 | 1.76 | 220 | — |
| BC-1 | 0.29 | 0.29 | 3.0 | 0.64 | 1.76 | 1.76 | — | -39.9 |
| Linac-2 | 0.29 | 2.13 | 0.64 | 0.64 | 1.76 | 0.71 | -35.85 | — |
| BC-2 | 2.13 | 2.13 | 0.64 | 0.23 | 0.71 | 0.71 | — | -17.2 |
| Linac-3 | 2.13 | 12.23 | 0.23 | 0.23 | 0.71 | 0.19 | +12.16 | — |

3. Undulator branch

The undulator branch is made up of 50 individual undulator [9] segments. Each segment has an in-vacuum permanent-magnet planar hybrid undulator with period length 18mm and nominal gap 6mm. The gap will be adjusted to yield effective undulator parameter K from 2.03 to 1.3. The permanent magnet is produced by rare-earth material $Sm_2Co_{17}$ with remanent field 1.1 Tesla. In order to provide good electrical boundaries, a 100μm thick copper-nickel coated foil covers the magnets. As shown in Tab.5 is the main parameters of the undulator magnet.

Tab.5. The main parameters of the undulator magnet

| parameter | value | |
|---|---|---|
| Type | Hybrid, in-vacuum | |
| Undulator period $\lambda_u$ (mm) | 18 | |
| Remanent field $B_r$ (T) | 1.1 | |
| Adjustable gap range (mm) | 3.5 ~ 6 | |
| Undulator parameter K | 2.03 | 1.3 |
| Undulator gap (mm) | 3.5 | 6 |
| Peak field (T) | 1.2 | 0.77 |

The undulator branch adopts FODO lattice, and only one undulator segment placed between two focusing quadrupoles. Such a short betatron period structure is advantageous in reducing the beam size. In order to achieve ultra hard 42keV XFEL radiation with relatively low energy electron beam, in-vacuum undulator is using. More detailed parameters are listed in Tab.6.

Tab.6. Parameters of the undulator branch

| parameter | value |
|---|---|
| Beam energy (GeV) | 12 |
| RMS normalized slice emittance (μm) | 0.25 |
| RMS energy spread (%) | 0.015 |
| Beam peak current (kA) | 3 |
| RMS beam length (fs) | 23 |
| Undulator period $\lambda_u$ (mm) | 18 |
| Undulator parameter K | 1.3 |
| Nominal undulator gap (mm) | 6 |
| Undulator segment length (m) | 3.456 |
| Number of segments | 50 |
| Focusing method | FODO |
| Supercell length (m) | 8.064 |
| Branch length (m) | 201.6 |
| Betatron phase advance per cell | 60° |
| Average beta function (m) | 7.64 |
| Radiation wavelength (nm) | 0.03 |
| Photon energy (keV) | 42 |
| Average radiation power (GW) | 21 |
| Bandwidth (%) | 0.08 |
| Photons per pulse | $2.1 \times 10^{11}$ |

Generally the tapering technique rises photon yield sharply because of the energy-loss compensation along the undulator, but it increases the undulator branch length and cost at the same time. As shown in Fig.8 is the results from GENESIS [10] simulation which combining with in-vacuum and tapering techniques, it is clear that a 200m undulator length is a good choice.

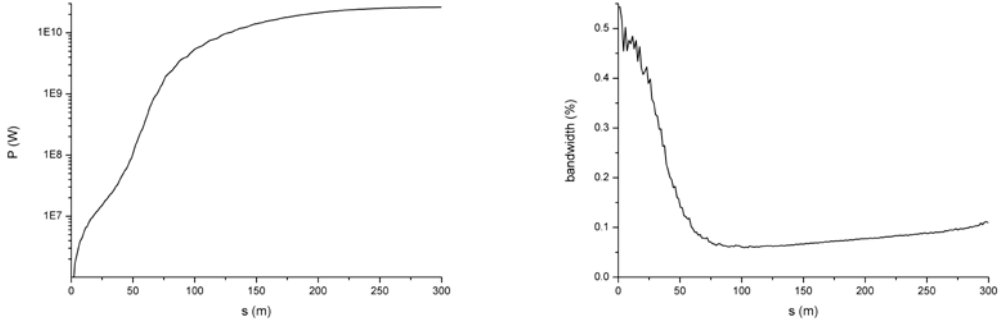

Fig.8. Radiation power (left) and bandwidth (right) along undulator branch

4. eRad beamline

High energy electron radiography technique depends on the use of particular amplifying magnetic lens to compensate for the small angel multiple Coulomb scattering (MCS) that occurs as the energetic electron pass through the sample. The use of such a magnetic lens turns the otherwise troubling complications of MCS into an asset. Electrons undergo the combined processes of small angle Coulomb scattering and ionization energy loss, each with its own unique dependence on material properties. These effects make possible the simultaneous determination of both material amounts and material identification.

The eRad requires a high energy beam to penetrate sample while keeping the deflection angle and energy loss small enough to allow good spatial resolution. To improve the resolution, the image blur should be small as possible. It is typically dominated by chromatic aberrations due to energy spread of the injected beam in combination with the spread of energy loss after passing through the sample. The image blur can be expressed as

$$\Delta x = T_{126}\theta\delta/M \ . \tag{1}$$

In which $\theta$ is the deflection angle mainly due to multiple Coulomb scattering (MCS) and $\delta = \Delta E/E_0$ is the relative energy deviation, $E_0$ is the energy of reference particle, $T_{126}$ (for x direction) and $T_{346}$ (for y direction) represents the second order chromatic aberration coefficients which decided by the lens system, $M$ is the image amplifying ratio.

The 1nC eRad electron bunch travels a long trip (~1km) in a separate branch adjacent to the XFEL before reaching the sample. At the same time, the transverse beam size expands 5~6 times by increasing the beta function to 1000m, and this can help illuminating ~1mm view field and decreasing the image blur caused by the beam's transverse insufficient modulation [11]. In the case of eRad bunch described in Tab.1, the rms effective deflection angle due to transverse insufficient modulation is equal to $\sqrt{\varepsilon/\beta} = 0.22$μrad.

As shown in Fig.9 is the schematic of 12GeV eRad beamline and its twiss parameters, the totally length is about 50m, the second order chromatic aberration coefficients $T_{126}$ and $T_{346}$ are 107m and 59m separately.

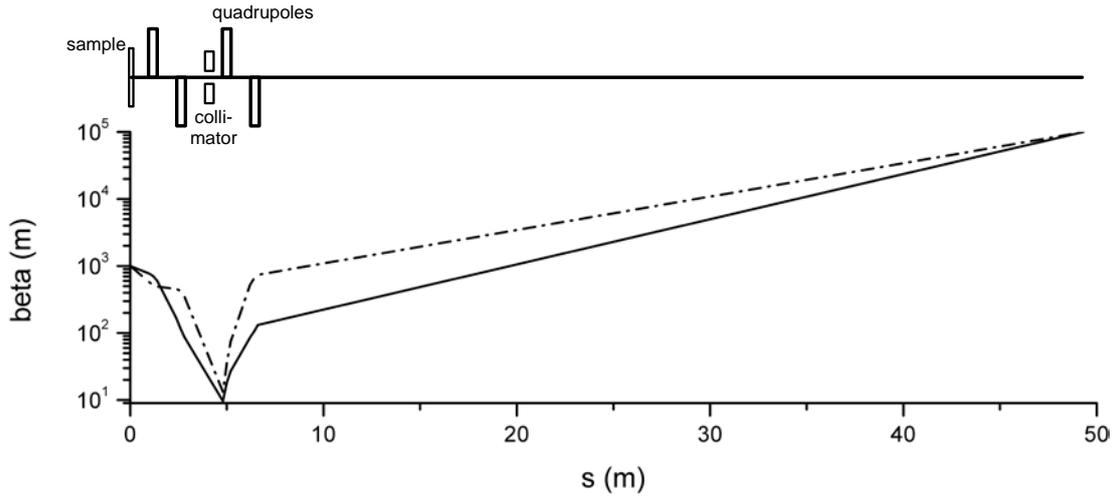

Fig.9. Schematic of 12GeV eRad beamline and its twiss parameters (when *M*=10), the solid line represents $\beta_x$ and the dash line represents $\beta_y$.

What was shown in Fig.10 is the simulation results by Monte Carlo code GEANT4 [12]. In simulation, we choose the copper sample 0.1mm long, the rms plane projection deflection angle is 117μrad, and the average energy loss is about 0.12MeV which lead to $1\times10^{-5}$ additional energy spread. In contrast, ionization energy loss can be ignored for the much bigger incident beam energy spread 0.2%. Therefore, the image blur can be calculated as 2.5μm for x direction and 1.4μm for y direction. In fact, the angle sort function can be carried out by the collimator on the Fourier plane, the electrons with greater angle can be cut off and higher resolution ability can be achieved.

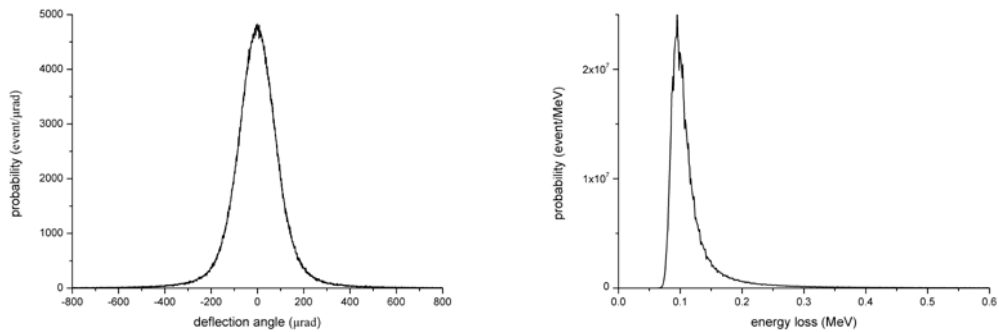

Fig.10. The distribution of deflection angle and energy loss simulated by Geant4 when 12GeV electron beam pass through the 0.1mm long copper sample

5. Conclusion and discussion

In the normal conductive accelerator scenario described in this article, the XFEL electron beam and eRad electron beam can be accelerated within the same operation mode. All machine parameters remain unchanged except the timing trigger of injector laser system which can be easily achieved online. As the two kinds of electron bunches undergo different branches after the linac and the optical path difference is tens nanoseconds, it is possible to separate them in the switch yard by kicker and septum magnet.

The transverse emittance of XFEL electron beam plays a key role in the photon yield. What was shown in Fig.11 is the time-dependent simulation results by GENESIS, normalized slice emmitance 0.3μm is the minimum requirement. Except CSR and space charge effect mentioned above, the influence of wake field on the transverse emittance cannot be ignored.

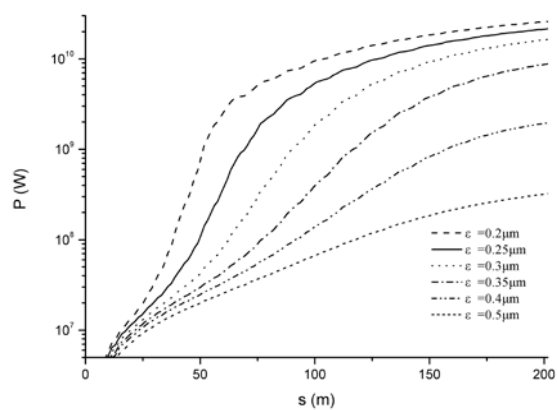

Fig.11. The radiation power of 42keV photons along undulator while the normalized emittance is different

The imaging principle of high energy eRad is originated in high energy proton radiography, but the specific implementation effectiveness has yet to be practice. The purpose of amplifying beamline is to reduce the impact of point spread function (PSF) in the scintillator screen. The improvement of incident beam energy spread can further upgrade the spatial resolution of eRad, especially for a relatively thin sample.

In fact, the dynamic studies require multiple pulses which haven't been mentioned in this article. To achieve high enough beam quality for multiple pulse XFEL electron bunches and eRad electron bunches, some research are under study in which the focus is beam quality degradation caused by wake field and beam loading.